# Ferroelectric Switching Pathways and Domain Structure of SrBi$_2$(Ta,Nb)$_2$O$_9$ from First Principles

*Nabaraj Pokhrel   Elizabeth A. Nowadnick*

Nabaraj Pokhrel
Department of Physics, University of California, Merced, Merced, California 95343, USA
Email Address: npokhrel@ucmerced.edu
Prof. Elizabeth A. Nowadnick
Department of Materials Science and Engineering, University of California, Merced, Merced, California 95343, USA
Email Address: enowadnick@ucmerced.edu

Several families of layered perovskite oxide ferroelectrics exhibit a coupling between polarization and structural order parameters, such as octahedral rotation distortions. This coupling provides opportunities for novel electric field-based manipulation of material properties, and also stabilizes complex domain patterns and domain wall vortices. Amongst layered perovskites with such coupled orders, the Aurivillius-phase oxides SrBi$_2$B$_2$O$_9$ (B=Ta, Nb) are well-known for their excellent room temperature ferroelectric performance. This work combines group theoretic analysis with density functional theory calculations to examine the ferroelectric switching processes of SrBi$_2$B$_2$O$_9$. Low-energy two-step ferroelectric switching paths are identified, with polarization reversal facilitated by structural order parameter rotations. Analysis of the domain structure reveals how the relative energetics of the coupled order parameters translates into a network of several distinct domain wall types linked by domain wall vortex structures. Comparisons are made between the ferroelectric switching and domain structure of SrBi$_2$B$_2$O$_9$ and those of the layered $n$=2 Ruddlesden-Popper hybrid improper ferroelectrics. The results provide new insight into how ferroelectric properties may be optimized by engineering the complex crystal structures of Aurivllius-phase oxides.

## 1 Introduction

The Aurivillius phases are a family of layered oxides with general chemical formula Bi$_{2m}$A$_{n-m}$B$_n$O$_{3(n+m)}$ and a crystal structure composed of perovskite-like slabs interleaved with Bi$_2$O$_2$ layers [1]. Several Aurivillius-phase oxides are ferroelectrics which are well known for their fatigue resistance, low coercive fields, and potential application in ferroelectric random access memory [2, 3, 4, 3, 5, 6]. Among these, SrBi$_2$Ta$_2$O$_9$ and SrBi$_2$Nb$_2$O$_9$ are well characterized ($m$=1, $n$=2) examples, which both have moderate electrical polarizations of ≈10 $\mu$C cm$^{-2}$ and low coercive fields of ≈35 kVcm$^{-1}$ [7, 8]. These materials are isostructural and crystallize in the polar $A2_1am$ space group at room temperature.

At high temperature, SrBi$_2$Ta$_2$O$_9$ and SrBi$_2$Nb$_2$O$_9$ crystallize in the high-symmetry $I4/mmm$ structure, shown in Figure 1(a). The room temperature polar $A2_1am$ structure, shown in Figure 1(b), can be decomposed into three structural distortions that transform like irreducible representations (irreps) of $I4/mmm$: a polar distortion that transforms like $\Gamma_5^-$ and two octahedral rotation distortions that transform like $X_2^+$ and $X_3^-$ (Figure 1(c-e)). A Landau expansion of the free energy about $I4/mmm$ allows a trilinear coupling between these three structural distortions:

$$F_{\text{tri}} = \alpha Q_{X_3^-} Q_{X_2^+} Q_{\Gamma_5^-}. \tag{1}$$

SrBi$_2$Ta$_2$O$_9$ and SrBi$_2$Nb$_2$O$_9$ exhibit distinct temperature-dependent structural phase transition sequences. As temperature lowers, SrBi$_2$Ta$_2$O$_9$ undergoes a sequence of two phase transitions: at ≈820 K, the $X_3^-$ octahedral tilt condenses leading to a structure with $Amam$ symmetry, and then at ≈650 K the polar and $X_2^+$ distortions condense and establish the polar $A2_1am$ structure [9, 10]. In contrast, SrBi$_2$Nb$_2$O$_9$ undergoes a direct avalanche-type transition from $I4/mmm$ to $A2_1am$ at ≈830K, where all three structural distortions condense together [9, 11]. The trilinear coupling in Equation 1 has been used to understand these phase transition sequences, and in particular the avalanche-type transition in SrBi$_2$Nb$_2$O$_9$ has been attributed to a stronger trilinear coupling coefficient $\alpha$ in this material [2, 12, 13].

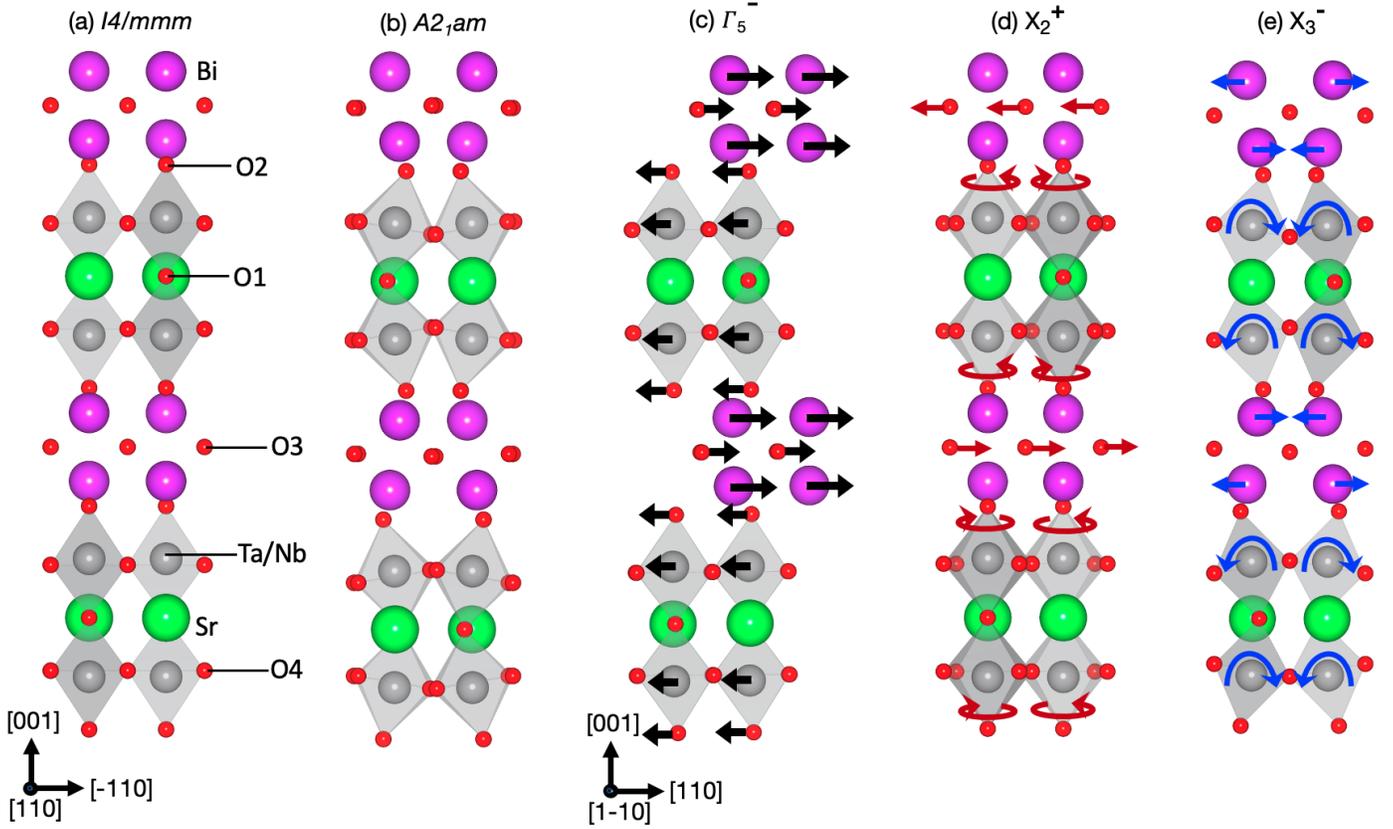

Figure 1: SrBi$_2$(Ta,Nb)O$_9$ crystal structure: (a) high-symmetry tetragonal $I4/mmm$ structure, and (b) polar orthorhombic $A2_1am$ structure. The distorted $A2_1am$ structure can be decomposed into three structural distortions that transform like irreducible representations of $I4/mmm$: (c) a polar distortion that transforms like $\Gamma_5^-$, (d) an octahedral rotation that transforms like $X_2^+$, (e) an octahedral tilt that transforms like $X_3^-$. The axes (in a tetragonal setting) shown under panel (a) are used for all panels except (c).

The trilinear coupling in Equation 1 also has important implications for the ferroelectric switching processes in SrBi$_2$(Ta,Nb)$_2$O$_9$. Namely, when the polarization reverses ($Q_{\Gamma_5^-} \to -Q_{\Gamma_5^-}$) one of the two octahedral rotations must also reverse, so that Equation 1 remains invariant. This means that the energetics of the octahedral rotation distortions play an important role in determining the ferroelectric switching mechanism. However, the role of the trilinear coupling in the SrBi$_2$(Ta,Nb)$_2$O$_9$ ferroelectric switching process has received minimal attention so far. Although a full description of the dynamic ferroelectric switching process is challenging, simpler approaches based on calculating *intrinsic* energy barriers for coherent reversal of a single infinite domain have proved insightful when applied to other ferroelectrics [14, 15, 16, 17]. In this work, we use a combination of density functional theory (DFT) calculations and group theoretic analysis to explore and enumerate intrinsic ferroelectric switching paths for SrBi$_2$(Ta,Nb)$_2$O$_9$, and understand their connection to the domain structure and domain wall-based switching.

Ferroelectricity characterized by a trilinear coupling between octahedral rotations and polarization is not just a feature of $n=2$ Aurivillius oxides, but occurs more generally in layered oxide ferroelectrics [18], such as the $n=2$ Ruddlesden-Popper oxides [19], $n=2$ Dion-Jacobson phases [20], and perovskite superlattices [21]. Previous work on ferroelectric switching in $n=2$ Ruddlesden-Popper Ca$_3$Ti$_2$O$_7$ and Sr$_3$Sn$_2$O$_7$ has revealed that the trilinear coupling in these materials leads to several possible symmetry-distinct switching paths, with the lowest energy possibility being a two-step path that passes through an intermediate antipolar structure [16, 22, 23]. The trilinear coupling also is encoded in the domain structure of $Ca_3Ti_2O_7$ as a complex network of meandering domain walls and domain wall vortices where three domain walls merge, as well as switching kinetics involving the creation of domain wall vortex/anti-vortex pairs at structural antiphase boundaries [24]. Interestingly, previous experimental work has revealed that $SrBi_2Ta_2O_9$ exhibits a strikingly similar domain structure and switching kinetics [25, 26, 27], which we further explore in this work.

Table 1: Experimental and DFT lattice parameters (in Å) for $SrBi_2Ta_2O_9$ and $SrBi_2Nb_2O_9$ in the $I4/mmm$ and $A2_1am$ structures.

|  | Lattice parameter | $SrBi_2Ta_2O_9$ DFT | $SrBi_2Ta_2O_9$ Expt.[28] | $SrBi_2Nb_2O_9$ DFT | $SrBi_2Nb_2O_9$ Expt.[29] |
|---|---|---|---|---|---|
| $I4/mmm$ | $a$ | 3.91 | 3.917 | 3.89 | - |
|  | $c$ | 24.47 | 25.114 | 24.47 | - |
| $A2_1am$ | $a$ | 5.53 | 5.52 | 5.54 | 5.519 |
|  | $b$ | 5.54 | 5.52 | 5.54 | 5.515 |
|  | $c$ | 24.95 | 25.02 | 24.95 | 25.112 |

## 2 Results

### 2.1 Ground state crystal structure

To set the stage for our work, we start by performing DFT structural relaxations of $SrBi_2Ta_2O_9$ and $SrBi_2Nb_2O_9$ in their polar $A2_1am$ and high-symmetry $I4/mmm$ structures. Table 1 compares our DFT-computed lattice parameters with their experimental values, and shows good agreement for both compounds. We next decompose the DFT and experimental $A2_1am$ structures with respect to the high-symmetry $I4/mmm$ reference structure, which reveals the amplitudes of the $\Gamma_5^-$, $X_2^+$, and $X_3^-$ distortions (Figure 1c-e). Table 2 shows the result of this decomposition. The distortion amplitudes for $SrBi_2Ta_2O_9$ and $SrBi_2Nb_2O_9$ are quite similar, which is expected given that Ta and Nb have very similar ionic radii.

The decomposition in Table 2 shows that the $X_3^-$ octahedral tilt has the largest amplitude and involves a significant deformation of the $BO_6$ octahedra: the apical oxygen in the SrO layer (O1) displaces roughly half the amount as the apical oxygen bordering the $Bi_2O_2$ layer (O2), due to the larger size of Sr compared to Bi. Bi displacements also make a fairly large contribution to this distortion. The $\Gamma_5^-$ polar distortion has the next largest amplitude, and arises from a displacement of the $Bi_2O_2$ layers against the perovskite blocks. Finally, the $X_2^+$ amplitude is the smallest, and is primarily composed of an $a^0a^0c^+$ octahedral rotation with a smaller contribution from the $Bi_2O_2$-layer O3 displacements. The $X_2^+$ distortion amplitude from DFT shows excellent agreement with the experimental value, whereas the $X_3^-$ and $\Gamma_5^-$ amplitudes from DFT are larger than those from experiment. For the $X_3^-$ distortion, this is because the oxygen displacements are significantly smaller in the experimental structure. Similarly, DFT overestimates the amplitudes of Bi and O3 displacements in the $\Gamma_5^-$ distortion.

### 2.2 Order parameters and metastable structural phases

Having analyzed the ground state structure, we next introduce order parameters to represent the structural distortions described above. These order parameters provide a framework for us to systematically explore a family of metastable structural phases that are relevant for constructing ferroelectric switching pathways in the next section.

Table 2: Decomposition of the DFT and experimental $A2_1am$ structures with respect to the high symmetry reference structure $I4/mmm$. The distortion amplitudes are reported in Å for a 56 atom unit cell. The experimental structures for SrBi$_2$Ta$_2$O$_9$ and SrBi$_2$Nb$_2$O$_9$ are taken from Refs. [28] and [29], respectively. $O4_\parallel$ and $O4_\perp$ are displacements along [110] and [1-10], respectively.

| Atom | SrBi$_2$Ta$_2$O$_9$ | | | | | | SrBi$_2$Nb$_2$O$_9$ | | | | | |
|---|---|---|---|---|---|---|---|---|---|---|---|---|
| | $\Gamma_5^-$ | | $X_2^+$ | | $X_3^-$ | | $\Gamma_5^-$ | | $X_2^+$ | | $X_3^-$ | |
| | DFT | Expt. | DFT | Expt. | DFT | Expt. | DFT | Expt. | DFT | Expt. | DFT | Expt. |
| Sr | -0.06 | -0.07 | - | - | -0.01 | -0.07 | -0.02 | -0.24 | - | - | -0.01 | -0.05 |
| Ta/Nb | 0.21 | 0.07 | - | - | -0.03 | -0.03 | 0.09 | -0.31 | - | - | -0.03 | -0.03 |
| Bi | -1.06 | -0.67 | - | - | 0.48 | 0.41 | -1.07 | -0.47 | - | - | 0.47 | 0.24 |
| O1 | 0.29 | 0.21 | - | - | 0.53 | 0.43 | 0.34 | 0.18 | - | - | 0.55 | 0.42 |
| O2 | 0.44 | 0.25 | - | - | -1.14 | -0.80 | 0.50 | 0.13 | - | - | -1.16 | -0.75 |
| O3 | -0.64 | -0.28 | 0.18 | 0.12 | -0.12 | -0.05 | -0.63 | -0.15 | 0.19 | 0.08 | -0.12 | 0.00 |
| $O4_\parallel$ | 0.45 | 0.31 | -0.52 | -0.53 | -0.89 | -0.70 | 0.52 | 0.49 | -0.69 | -0.88 | -0.91 | -0.70 |
| $O4_\perp$ | 0.33 | 0.22 | 0.01 | 0.02 | - | - | 0.37 | 0.35 | 0.03 | 0.01 | - | - |
| Total: | 1.48 | 0.88 | 0.55 | 0.54 | 1.63 | 1.21 | 1.52 | 0.90 | 0.72 | 0.88 | 1.65 | 1.14 |

The $\Gamma_5^-$, $X_2^+$, and $X_3^-$ distortions are described by two-dimensional order parameters, that is, each distortion is characterized by an amplitude $Q$ and a phase $\phi$. The two-dimensional order parameter for each distortion $i=(\Gamma_5^-, X_2^+, X_3^-)$ can be written as [30, 16]:

$$\boldsymbol{\eta}^i = (\eta_1^i, \eta_2^i) = Q_i e^{i\phi_i}. \tag{2}$$

With two-dimensional order parameters, the trilinear coupling in Equation 1 can be written as

$$F_{\text{tri}} = \alpha[\eta_1^{X_3^-} \eta_1^{X_2^+} (\eta_1^{\Gamma_5^-} + \eta_2^{\Gamma_5^-}) + \eta_2^{X_3^-} \eta_2^{X_2^+} (\eta_1^{\Gamma_5^-} - \eta_2^{\Gamma_5^-})]. \tag{3}$$

There are three symmetry distinct choices for the direction of each two-dimensional order parameter: $(a,0)$, $(a,a)$, and $(a,b)$, where $a$ and $b$ are real numbers. Each choice defines a structure with a different symmetry (formally, a different isotropy subgroup of $I4/mmm$). For example, the $(a,0)$ direction of $X_3^-$ defines an $Amam$-symmetry structure with an $a^-a^-c^0$ octahedral tilt pattern (in Glazer notation), as shown in Figure 1(e). The $(a,a)$ direction of $X_3^-$ gives a structure with symmetry $P4_2/mnm$, which has $a^-b^0b^0$ and $b^0a^-b^0$ octahedral tilts in alternating perovskite layers (Figure 2(a)). Table 3 reports the subgroups of $I4/mmm$ generated by the $(a,0)$ and $(a,a)$ directions of the $\Gamma_5^-$, $X_2^+$, and $X_3^-$ irreps, as well as those established by all combinations of these irreps. We note that for some combinations of two irreps, the third irrep is induced (that is, it does not further lower the symmetry).

We next perform DFT structural relaxations of SrBi$_2$Ta$_2$O$_9$ and SrBi$_2$Nb$_2$O$_9$ with symmetry constrained to each of the subgroups in Table 3, and report the resulting energies above the $A2_1am$ ground state (see the Supporting Information for lattice parameters and distortion amplitudes). We find that the lowest energy metastable phase is $P4_2/mnm$, which is 22.6 (41.3) meV per formula unit ((f.u.)$^{-1}$) above the $A2_1am$ ground state for SrBi$_2$Ta$_2$O$_9$ (SrBi$_2$Nb$_2$O$_9$). The second lowest energy phase has symmetry $Pnam$, with energy 38.8 (41.9) meV(f.u.)$^{-1}$ above the ground state.

The $Pnam$ structure (Figure 2(b)) is closely related to the polar $A2_1am$ structure: they both exhibit an $a^-a^-c^+$ octahedral rotation pattern, with the difference lying in the relative "sense" of the $X_2^+$ octahedral rotation across the Bi$_2$O$_2$ layer. This results in a trilinear coupling between the octahedral rotations and an "antipolar" distortion in $Pnam$ that transforms like the $M_5^-$ irrep. As shown in Figure 2(b), the $M_5^-$ distortion is characterized by the displacement of adjacent perovskite slabs against each other, as well as displacements of the layers of each Bi$_2$O$_2$ slab against each other. Significantly, if two domains of $A2_1am$ with opposite polarization direction are stacked along the long ($c$) axis, the antipolar $Pnam$ structure would naturally form at their interface. A similar feature, termed a "stacking" domain wall, has been shown to play an important role in the switching mechanism of the $n$=2

Table 3: Isotropy subgroups of $I4/mmm$ generated by distinct directions of the $\Gamma_5^-$, $X_2^+$, and $X_3^-$ order parameters and all combinations of two of these order parameters. For space groups where an antipolar $M_5^-$ order parameter also is allowed, its direction is given. The total energies of SrBi$_2$Ta$_2$O$_9$ and SrBi$_2$Nb$_2$O$_9$ after structural relaxations with symmetry confined to each space group are reported with respect to the energy of $A2_1am$. For $P2_1am$ and $P2_1nm$, where one of the order parameters is along a low symmetry $(a,b)$ direction, the reported energy is the maximum value obtained from a nudged elastic band calculation. For structural phases that relax to a higher symmetry space group, that space group is indicated in the energy column.

| Irrep | $\eta^{\Gamma_5^-}$ | $\eta^{X_2^+}$ | $\eta^{X_3^-}$ | $\eta^{M_5^-}$ | Space group | Energy (meV(f.u.)$^{-1}$) SrBi$_2$Ta$_2$O$_9$ | SrBi$_2$Nb$_2$O$_9$ |
|---|---|---|---|---|---|---|---|
| $\Gamma_5^-$ | $(a,0)$ | | | | $Imm2$ | 168.10 | 174.57 |
| | $(a,a)$ | | | | $F2mm$ | 160.39 | 174.62 |
| $X_2^+$ | | $(a,0)$ | | | $Acam$ | 272.41 | 284.22 |
| | | $(a,a)$ | | | $P4/mbm$ | 302.06 | 316.31 |
| $X_3^-$ | | | $(a,0)$ | | $Amam$ | 63.83 | 88.54 |
| | | | $(a,a)$ | | $P4_2/mnm$ | 22.60 | 41.33 |
| $X_2^+ \oplus X_3^-$ | | $(a,a)$ | $(c,0)$ | | $A2_1am$ | 0 | 0 |
| | | $(0,a)$ | $(b,0)$ | $(0,-c)$ | $Pnam$ | 38.86 | 41.89 |
| | | $(a,0)$ | $(b,b)$ | $(c,c)$ $(d,-d)$ | $C2mm$ | ($P4_2/mnm$) | |
| | | $(a,a)$ | $(b,c)$ | $(d,0)$ $(0,-e)$ | $P2_1am$ | 44.35 | 59.67 |
| | | $(a,a)$ | $(b,0)$ | $(c,d)$ $(e,0)$ | $P2_1nm$ | 31.97 | 50.29 |
| $\Gamma_5^- \oplus X_2^+$ | $(a,a)$ | $(0,b)$ | | | $Cm2a$ | 101.64 | 96.44 |
| $\Gamma_5^- \oplus X_3^-$ | $(a,a)$ | | $(0,b)$ | | $C2mm$ | ($Amam$) | |

Table 4: Decomposition of the DFT-relaxed $Pnam$ phase into symmetry adapted modes of $I4/mmm$ for SrBi$_2$Ta$_2$O$_9$ and SrBi$_2$Nb$_2$O$_9$. The distortion amplitudes are given in Å for a 56 atom computational cell.

| Atom | SrBi$_2$Ta$_2$O$_9$ | | | SrBi$_2$Nb$_2$O$_9$ | | |
|---|---|---|---|---|---|---|
| | $M_5^-$ | $X_2^+$ | $X_3^-$ | $M_5^-$ | $X_2^+$ | $X_3^-$ |
| Sr | 0.41 | - | -0.01 | 0.36 | - | -0.02 |
| Ta/Nb | -0.73 | - | -0.03 | -0.65 | - | -0.01 |
| Bi | -0.23 | - | 0.50 | -0.25 | - | 0.47 |
| O1 | -0.64 | - | 0.54 | -0.71 | - | 0.55 |
| O2 | -0.89 | - | -1.17 | -0.98 | - | -1.18 |
| O3 | 0.03 | 0.04 | -0.18 | 0.03 | 0.04 | -0.18 |
| O4$_\parallel$ | -0.97 | 0.58 | -0.92 | -1.07 | 0.76 | -0.93 |
| O4$_\perp$ | 0.84 | 0.01 | - | 0.91 | 0.03 | - |
| Total: | 1.91 | 0.58 | 1.67 | 2.02 | 0.75 | 1.67 |

Ruddlesden-Popper ferroelectrics [16, 22]. Table 4 shows the decomposition of the $Pnam$ structure with respect to $I4/mmm$. The $X_2^+$ and $X_3^-$ distortion amplitudes are similar to those of the polar $A2_1am$ structure in Table 2. The $M_5^-$ distortion has the largest amplitude, with oxygen displacements make the largest contribution and significant contributions also coming from the cations, in particular Ta/Nb. The Bi displacements are quite small compared to those present in the $\Gamma_5^-$ distortion. We note that, as indicated in Table 3, the $M_5^-$ distortion is allowed by symmetry in several other low-symmetry structures besides $Pnam$.

## 2.3 Ferroelectric switching pathways

In this section we make use of the family of metastable structural phases introduced above to systematically enumerate possible ferroelectric switching pathways and explore their energy barriers. The trilinear coupling between the polarization and octahedral rotations from Equation 3 requires that when the polarization reverses ($\eta^{\Gamma_5^-} \longrightarrow -\eta^{\Gamma_5^-}$), one but not both of the octahedral rotation order parameters must reverse. This can be accomplished by turning off the polarization and one of the octahedral rotation order parameters and then turning them on again pointing in the opposite direction, which leads to a "one-step" switching path. Alternatively, polarization reversal can be achieved by "rotating" the two-dimensional polar and/or octahedral rotation order parameters (changing $\phi$) while keeping the amplitude $Q$ finite. These order parameter rotations will generally result in "two-step" switching paths, where the barrier (highest energy) structure is visited twice.

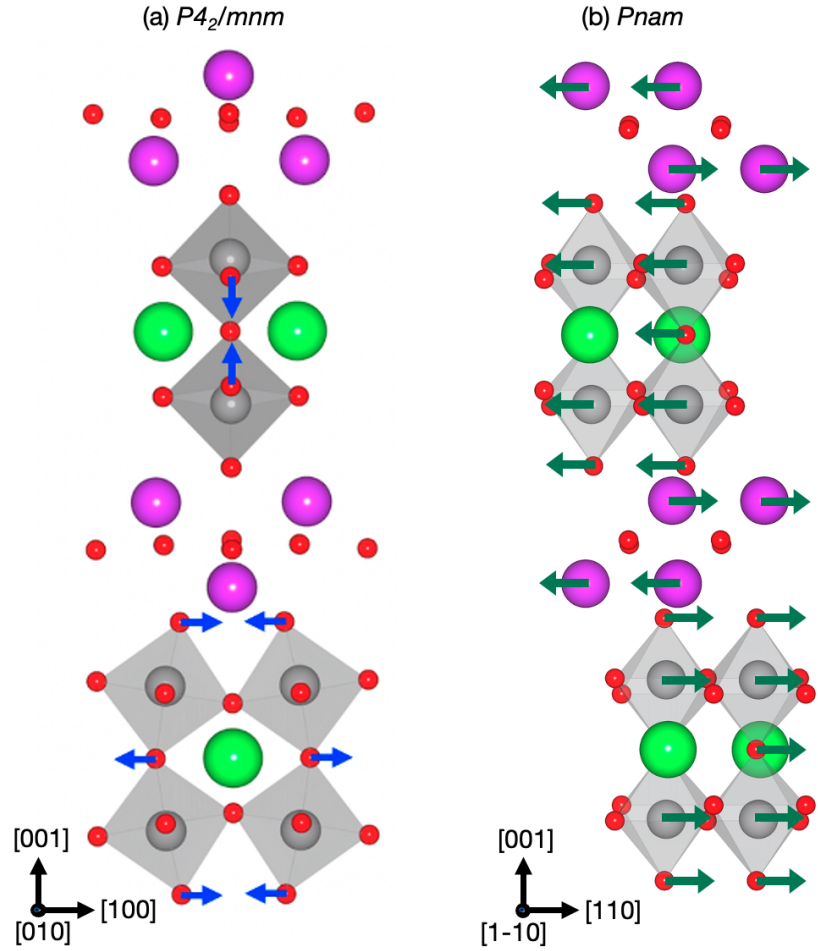

Figure 2: Low-energy metastable structural phases: (a) $P4_2/mnm$, which is characterized by out-of-phase octahedral tilts about [100] and [010] in adjacent perovskite layers (blue arrows), and (b) $Pnam$, which hosts $X_3^-$ and $X_2^+$ distortions that couple to an antipolar distortion that transforms like irrep $M_5^-$ (green arrows).

We first consider ferroelectric switching paths where the $X_2^+$ order parameter reverses along with the polarization. Figure 3 summarizes these paths, obtained with nudged elastic band (NEB) calculations, for SrBi$_2$Ta$_2$O$_9$ (images of how the crystal structure evolves, as well as results for SrBi$_2$Nb$_2$O$_9$, are shown in the Supporting Information). In the one-step path shown in Figure 3(a), the amplitudes of the polar and $X_2^+$ order parameters smoothly go to zero upon approaching the barrier structure with symmetry $Amam$. The total energy as a function of switching coordinate is shown in Figure 3(d); the $Amam$ barrier energy is 64 meV(f.u.)$^{-1}$.

Figure 3(b) shows an alternative path which takes advantage of the two-dimensional nature of the order parameters: instead of keeping the $X_3^-$ order parameter oriented along $(a,0)$ during the entire switching process, $X_3^-$ can instead rotate to lie along $(a,a)$ when $Q_{X_2^+} = Q_{\Gamma_5^-} = 0$, thus establishing the $P4_2/mnm$ phase. Then, as the $X_2^+$ and $\Gamma_5^-$ order parameters turn on pointing in the opposite direction, the $X_3^-$ order parameter rotates back to its original $(a,0)$ direction. At intermediate points between the $A2_1am$ and $P4_2/mnm$ structures, the $X_3^-$ order parameter lies along the low-symmetry $(a,b)$ direction, establishing a $P2_1nm$ structure. Traversing the path from $A2_1am$ to $P4_2/mnm$, the $X_2^+$ and $\Gamma_5^-$ amplitudes go smoothly to zero, while a small-amplitude $M_5^-$ distortion (symmetry-allowed in $P2_1nm$) turns on and then off again. Figure 3(d) show that this is a two-step switching path, with a $P2_1nm$-symmetry barrier of energy 32 meV(f.u.)$^{-1}$ visited twice during the switching process. This $P2_1nm$ barrier energy is about half the value of the one-step $Amam$ barrier, indicating that two-step switching is energetically favorable.

Next, in Figure 3(c) we consider reversing the $X_2^+$ order parameter by changing its phase $\phi$ by 180° (while maintaining a finite $Q_{X_2^+}$ throughout). At the midpoint of this path when $\Delta\phi = 90°$, the structure has symmetry $Pnam$. Between $A2_1am$ and $Pnam$, the $X_2^+$ order parameter lies along the low-symmetry $(a,b)$ direction, establishing a $P2_1am$-symmetry structure. The amplitudes of the $X_3^-$ and $X_2^+$ order parameters stay approximately constant throughout the switching process, whereas the $\Gamma_5^-$ polar distortion smoothly decreases going from $A2_1am$ to $Pnam$, whereas the $M_5^-$ distortion smoothly increases. The $P2_1am$-symmetry barrier of this two-step switching path is 44 meV(f.u.)$^{-1}$. Interestingly, close inspection of Figure 3(d) reveals that the $P2_1am$ barrier - the highest energy point along the switching path - does not occur at the midpoint of each step (switching coordinate values of 0.25 and 0.75), but occurs closer to $Pnam$. This occurs because the $\Gamma_5^-$ distortion is significantly more energy lowering than the $M_5^-$ distortion; we explore this point further in the Supporting Information.

Besides the switching paths discussed so far, there are a few other symmetry-distinct paths to consider for a complete enumeration. The first is the analogous one-step path to that in Figure 3(a), except where now the $X_3^-$ order parameter reverses by turning off/on. This results in a $Acam$-symmetry barrier with an energy of 272 meV(f.u.)$^{-1}$, see Table 3. In addition, switching may proceed via polarization rotation (changing $\phi$). This requires the $\Gamma_5^-$ order parameter to pass through the $(a,0)$ direction. The only space groups in Table 3 with $\Gamma_5^-$ along $(a,0)$ are $Imm2$ and $C2mm$. Polarization switching via $Imm2$ proceeds by making two 90° rotations of the $\Gamma_5^-$ order parameter, where at each step the $X_3^-$ and $X_2^+$ amplitudes go to zero at the $Imm2$ barrier, yielding a barrier energy of 168 meV(f.u.)$^{-1}$. Polarization reversal via $C2mm$ involves rotating all three order parameters. However, our calculations show that this path relaxes to the Figure 3(c) path. Thus, the lowest energy switching paths for SrBi$_2$Ta$_2$O$_9$ are those shown in Figure 3. Calculations on SrBi$_2$Nb$_2$O$_9$ give qualitatively similar results, with slightly higher barriers of 89, 50, and 60 meV(f.u.)$^{-1}$ for the $Amam$, $P2_1nm$, and $P2_1am$ barrier

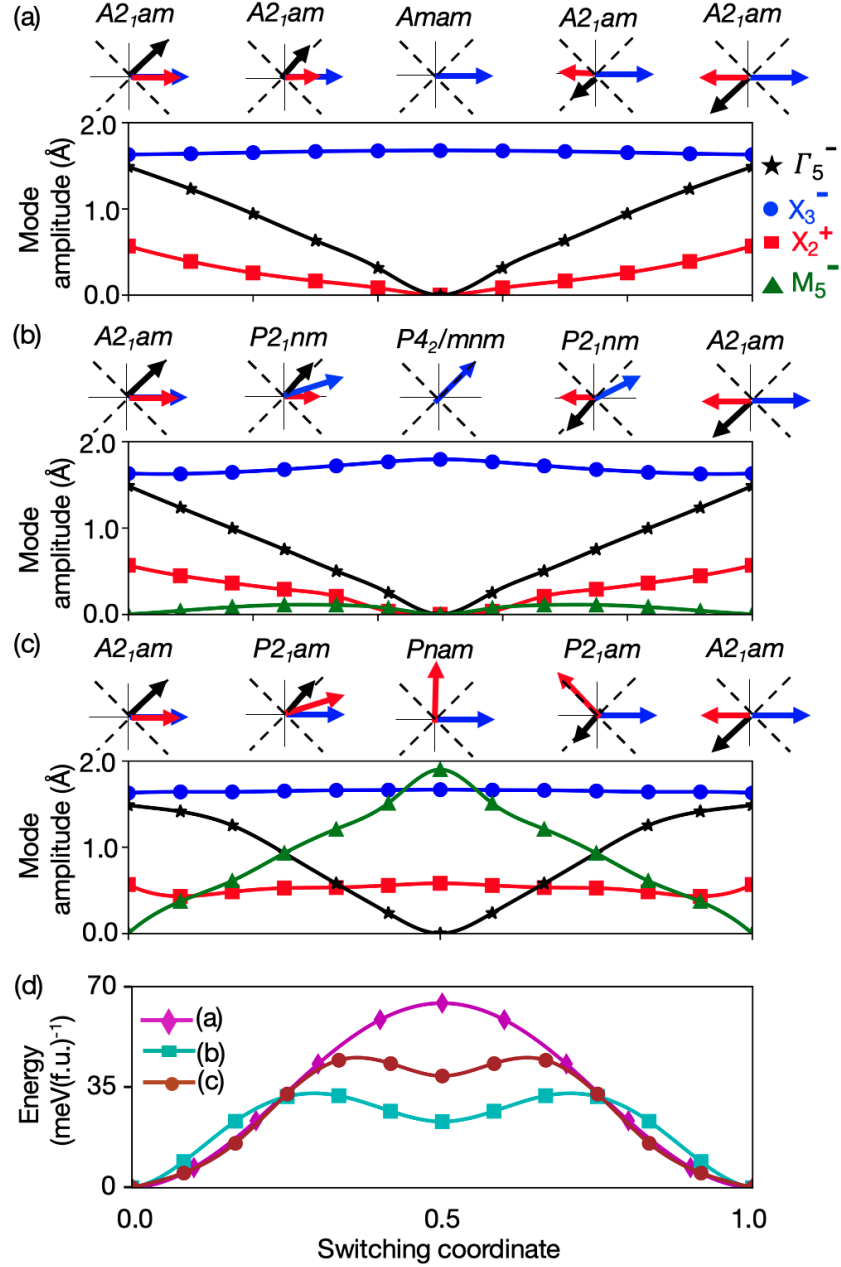

Figure 3: Ferroelectric switching paths that reverse the $X_2^+$ octahedral rotation for $SrBi_2Ta_2O_9$: (a) one-step switching, (b) two-step switching via $P4_2/mnm$, and (c) two-step switching via $Pnam$. The upper part of each panel shows how the $\Gamma_5^-$ (black), $X_2^+$ (red), and $X_3^-$ (blue) order parameters evolve during switching, and the bottom part shows the amplitudes as a function of switching coordinate, obtained from nudged elastic band calculations. (d) Total energy as a function of switching coordinate obtained from nudged elastic band calculations, for the paths shown in (a)-(c).

symmetries, respectively.

## 2.4 Domain structure

We next extend our analysis to investigate the domain structures of $SrBi_2Ta_2O_9$ and $SrBi_2Nb_2O_9$. We first enumerate the eight structural domains of the polar ground state. The $(a,0)$ and $(0,a)$ directions of the two-dimensional structural order parameters define orthorhombic twins of the same isotropy subgroup, with the setting of the orthorhombic space group ($A2_1am$ vs. $Bb2_1m$) distinguishing between the two twins. Figure 4 shows the eight structural domains, divided between the two orthorhombic twins. The four structural domains within each twin are divided into pairs with the same polarization direction (but opposite $X_3^-$ and $X_2^+$ order parameter directions). Five distinct types of domain walls can form between these domains [25, 26, 27]: two types of 90° ferroelectric/ferroelastic walls between domains in opposite orthorhombic twins (head-to-head and head-to-tail walls), two types of 180° ferroelectric walls (one where $X_3^-$ reverses across the wall and one where $X_2^+$ reverses), and an antiphase boundary where the polarization is the same direction on either side of the boundary but both the $X_3^-$ and the $X_2^+$ order parameters reverse.

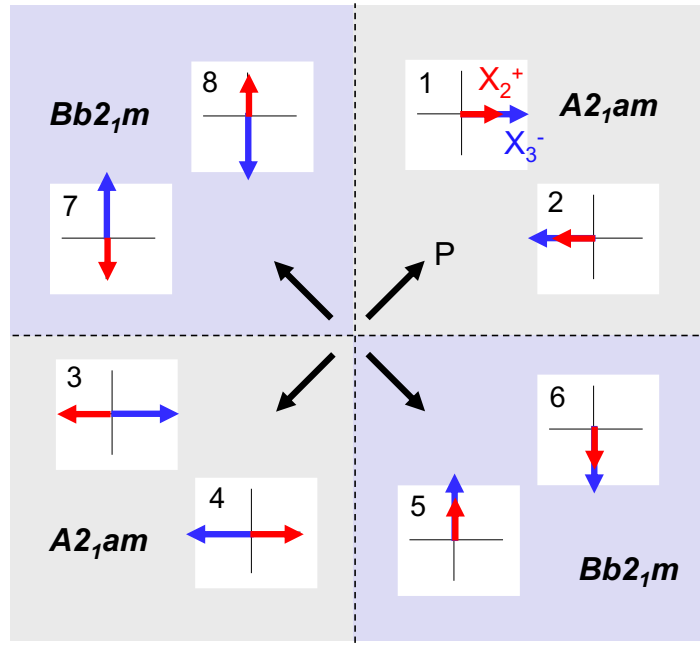

Figure 4: The eight structural domains of the polar ground state organized by orthorhombic twins (shaded grey and purple) and polarization direction.

Beyond domains and domain walls, in the last decade complex domain structures with topologically protected vortex-like features - where specific numbers of domain walls merge - have been uncovered in a range ferroelectric materials [31, 32, 33]. In addition to providing striking visualizations of topology in real space, these vortex structures impact domain nucleation and domain wall propagation which are key aspects of the ferroelectric switching process. Recently, three-fold domain wall vortices have been revealed in the $n=2$ Ruddlesden-Popper $Ca_{3-x}Sr_xTi_2O_7$ [24]. Motivated by the similar domain structure reported for $SrBi_2Ta_2O_9$, [25, 26, 27] we next assess the possible domain wall vortices that may form in $SrBi_2Ta_2O_9$ and $SrBi_2Nb_2O_9$.

Conveniently, we can use the intrinsic ferroelectric switching pathways constructed in the previous section to explore domain wall vortex structures. This is because intrinsic ferroelectric switching pathways, by definition, are energy trajectories through order parameter space between domains with opposite polarization directions. We can take a more generalized view to consider energy trajectories between any two domains (not just those with opposite polarization). In the past decades, this order parameter approach has been used to classify inhomogeneous structures in many systems with multi-component order parameters, for example improper ferroelectrics Gd$_2$(MoO$_4$)$_3$ [34, 35, 36] and hexagonal YMnO$_3$ [33]. In order parameter space, a domain wall corresponds to a trajectory between two domains, and a $n$-fold vortex corresponds to a closed path that passes through $n$ domains which starts and ends at the same domain. The energy hierarchy between different order parameter space trajectories determines the lowest energy closed path (vortex) that can be constructed.

| Path(s) | Domain wall type | Barrier symmetry | Barrier energy (meV(f.u.)$^{-1}$) | |
| --- | --- | --- | --- | --- |
| | | | SrBi$_2$Ta$_2$O$_9$ | SrBi$_2$Nb$_2$O$_9$ |
| domain 1→2 | Antiphase boundary | $Fmm2$ | 160.39 | 174.62 |
| domain 1→3 | 180° ferroelectric wall ($X_2^+$ reverses) | $P2_1nm$ | 31.97 | 50.29 |
| domain 1→4 | 180° ferroelectric wall ($X_3^-$ reverses) | $Acam$ | 272.41 | 284.22 |
| domain 1→(5,6,7,8) | 90° ferroelectric/ferroelastic wall | $P2_1nm$ | 31.97 | 50.29 |

Table 5: Summary of paths connecting domain 1 to all other structural domains in the polar $A2_1am$ structure and their interpretation as domain walls. The numbering of the domains is given in Figure 4. Note these energies also are given in Table 3, but are reproduced here to facilitate discussion of the domain structure.

We explore this energy hierarchy in SrBi$_2$Ta$_2$O$_9$ and SrBi$_2$Nb$_2$O$_9$ by starting with domain 1 in Figure 4 and constructing paths to each of domains 2-8, as summarized in Table 5. The path between domains 1 and 3 (180° ferroelectric wall with reversal of the $X_2^+$ order parameter) already was presented in Figure 3(b). Going from domain 1 to any of domains 5-8 corresponds to a 90° ferroelectric/ferroelastic wall. Interestingly, these paths can be derived from a small modification to the Figure 3(b) path: namely, in the second step the $X_3^-$ order parameter continues to rotate counterclockwise to the $(0,a)$ direction instead of returning to its original orientation (see Supporting Information for more details). Thus the only remaining paths to consider are those from domain 1 to domains 2 and 4. The domain 1 → 4 path is a 180° ferroelectric wall, where now the $X_3^-$ order parameter reverses along with the polarization. The domain 1 → 2 path is an antiphase boundary, where the polarization lies along the same direction in both domains but both the $X_2^+$ and $X_3^-$ order parameters are reversed. Both of these paths have high energy barriers because they require reversal of the large amplitude $X_3^-$ order parameter.

However, inspection of Figure 4 reveals that these high-energy domain 1 → (2,4) paths could be replaced by lower energy paths that involve going first from domain 1 to one of the domains in the orthorhombic twin (domains 5-8) via a $P2_1nm$-barrier path, and then traversing back to domain 2 or 4 via a second $P2_1nm$-barrier path. This results in a four-step path, where the low-energy $P2_1nm$ barrier structure is visited four times. This can be interpreted as the antiphase boundary being unstable to decaying into two 90° ferroelectric/ferroelastic walls, forming a three-fold domain wall vortex (shown schematically in Figure 5(a)). This provides understanding of the experimental observation [25, 26, 27] that antiphase boundaries form domain nucleation sites during switching, because an antiphase boundary splitting into two ferroelastic walls would nucleate a ferroelastic domain. Our calculations also predict that the 180° ferroelectric wall with $X_3^-$ reversal is unstable to decay into two ferroelastic walls, nucleating a ferroelastic domain in the same manner (Figure 5(b)).

## 3 Discussion

In this work we have explored the instrinsic ferroelectric switching pathways in Aurivillius-phase oxides SrBi$_2$Ta$_2$O$_9$ and SrBi$_2$Nb$_2$O$_9$ using group theoretic analysis together with DFT calculations. We find that two-step switching via a $P2_1nm$-symmetry barrier of 32 (50) meV(f.u.)$^{-1}$ for $SrBi_2Ta_2O_9$ ($SrBi_2Nb_2O_9$) is the lowest energy path, followed by two-step switching via a $P2_1am$ barrier of 44 (60) meV(f.u.)$^{-1}$. Both paths involve reversal of the $X_2^+$ octahedral rotation together with the polarization, so the larger barriers in $SrBi_2Nb_2O_9$ can be explained by the larger $X_2^+$ amplitude in this material (Table 2). Ferroelectric switching via reversal of the $X_3^-$ tilt together with the polarization has a significantly higher barrier, due to the larger amplitude of this distortion.

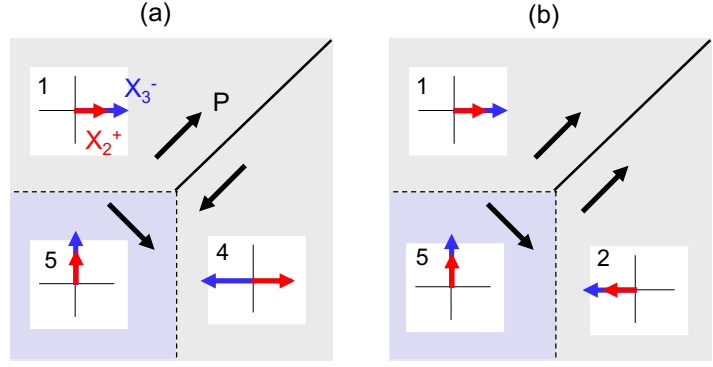

Figure 5: Three-fold domain wall vortices obtained by order parameter space energetics: (a) a three-fold vortex where a 180° ferroelectric wall merges with two ferroelastic walls, and (b) a three-fold vertex where an antiphase boundary merges with two ferroelastic walls. The black arrows indicate the polarization direction in each domain, the dashed lines indicate ferroelastic walls, and the grey/lavender color indicates opposite orthorhombic twins.

These computed intrinsic ferroelectric switching barriers are quite low and are comparable to those of prototypical ferroelectric perovskite oxides such as $BaTiO_3$ (20 meV(f.u.)$^{-1}$ [37]) and $PbTiO_3$ (50 meV(f.u.$^{-1}$) [14]), especially when the larger formula unit size of the Aurivillius oxides is taken into account. They also are similar to the computed barriers of the $n=2$ Ruddlesden-Popper hybrid improper ferroelectrics (e.g. 24 meV(f.u.)$^{-1}$ for $Sr_3Sn_2O_7$ [23], 64 meV(f.u.)$^{-1}$ for $Ca_3Ti_2O_7$ [16]). The low switching barriers of $SrBi_2Ta_2O_9$ and $SrBi_2Nb_2O_9$ are facilitated by the low-energy $P4_2/mnm$ phase. Further lowering the energy of the $P4_2/mnm$ phase, for example by judicious choice of $A$-site cation, may provide a useful strategy to optimize the switching barriers of these Aurivillius-phase oxides.

Our work also provides insight into the $SrBi_2B_2O_9$ domain structure. In particular, we find that it is energetically favorable for domain walls that involve reversal of the large amplitude $X_3^-$ tilt (180° ferroelectric walls and antiphase boundaries) to decay into two ferroelastic walls. This provides a natural explanation of the experimental observation of domain nucleation at antiphase boundaries and domain walls merging into groups of three (three-fold domain wall vortices). Furthermore, the $P2_1am$-barrier path describes switching via the motion of "stacking" domain walls along the long ($c$) axis, if such walls are present in the sample. Such "stacking" domain walls have been observed and play an important role ferroelectric switching in the $n=2$ Ruddlesden-Popper hybrid improper ferroelectrics [16, 23]. However, in the Ruddlesden-Popper materials the $Pnam$ antipolar structure is significantly closer in energy to the polar ground state (12 meV(f.u.)$^{-1}$ in $Ca_3Ti_2O_7$ [16], compared to ≈40 meV(f.u.)$^{-1}$ for $SrBi_2Ta_2O_9$ and $SrBi_2Nb_2O_9$), so stacking domain wall motion may be a less important switching mechanism for these Aurivillius oxides.

Taken together, our results provide new understanding of the mechanisms underlying the low coercive fields of Aurivillius oxides as well as their domain structure. We hope that this work stimulates new experimental investigations into this complex and technologically important class of ferroelectric oxides.

# 4 Computational Methods

We perform density functional theory (DFT) calculations using the Vienna *ab initio* Simulation Package (VASP) [38]. We use the PBEsol functional [39] and the following projector-augmented wave (PAW) VASP pseudopotentials: Sr_sv ($4s^2 4p^6 5s^2$), Bi ($6s^2 6p^3$), Ta ($6s^2 5d^3$), Nb_sv ($4s^2 4p^6 5s^1 4d^4$), and O_s ($2s^2 2p^4$). We employ a 600 eV plane-wave cutoff, a $6 \times 6 \times 2$ Monkhorst-Pack $k$-point mesh in a 56 atom computational cell, a $1 \times 10^{-7}$ eV energy convergence criterion, and a 2.5 meV Å$^{-1}$ force convergence criterion. For ferroelectric switching calculations, we use the nudged elastic band (NEB) method [40] to relax the intermediate structures along the switching paths. The ionic positions and lattice parameters are allowed to relax for each image in the NEB calculations, and the force convergence tolerance is increased to 10 meV Å$^{-1}$. Group theoretic analysis is performed with the aid of the ISOTROPY Software Suite [41, 42], and VESTA [43] is used to visualize crystal structures.

# Supporting Information

Supporting Information is available from the Wiley Online Library or from the author.

# Acknowledgements

This research was supported by the Office of Naval Research under Contract No. N00014-21-1-2957. We acknowledge use of computational resources provided by the Center for Functional Nanomaterials, which is a U.S. DOE Office of Science Facility, and the Scientific Data and Computing Center, a component of the Computational Science Initiative, at Brookhaven National Laboratory under Contract No. DE-SC0012704. We thank Sriram P. Ramkumar for useful discussions.

# Conflict of Interest

The authors declare no conflict of interest.

# Data Availability Statement

The data that support the findings of this study are provided in a Dryad repository [44].

# References

[1] B. Aurivillius, *Ark Kemi.* **1949**, *1* 463.

[2] J. M. Perez-Mato, M. Aroyo, A. García, P. Blaha, K. Schwarz, J. Schweifer, K. Parlinski, *Phys. Rev. B* **2004**, *70* 214111.

[3] C. A.-P. de Araujo, J. D. Cuchiaro, L. D. McMillan, M. C. Scott, J. F. Scott, *Nature* **1995**, *374*, 6523 627.

[4] Y. Shimakawa, Y. Kubo, Y. Tauchi, T. Kamiyama, H. Asano, F. Izumi, *Appl. Phys. Lett.* **2000**, *77*, 17 2749.

[5] M. H. Tang, Z. H. Sun, Y. C. Zhou, Y. Sugiyama, H. Ishiwara, *Appl. Phys. Lett.* **2009**, *94*, 21 212907.

[6] M. S. Bozgeyik, *Bull. Mater. Sci.* **2019**, *42*, 2 47.

[7] T. Mihara, H. Yoshimori, H. Watanabe, C. A. P. de Araujo, *Jpn. J. Appl. Phys.* **1995**, *34*, Part 1, No. 9B 5233.

[8] P. Yang, D. L. Carroll, J. Ballato, R. W. Schwartz, *J. Appl. Phys.* **2003**, *93*, 11 9226.

[9] P. Boullay, J. Tellier, D. Mercurio, M. Manier, F. Zuñiga, J. Perez-Mato, *Solid State Sci.* **2012**, *14*, 9 1367.

[10] A. Onodera, T. Kubo, K. Yoshio, S. Kojima, H. Yamashita, T. Takama, *Jpn. J. Appl. Phys.* **2000**, *39* 5711.

[11] A. Snedden, C. H. Hervoches, P. Lightfoot, *Phys. Rev. B* **2003**, *67* 092102.

[12] U. Petralanda, I. Etxebarria, *Phys. Rev. B* **2015**, *91* 184106.

[13] I. Etxebarria, J. M. Perez-Mato, P. Boullay, *Ferroelectrics* **2010**, *401*, 1 17.

[14] S. Beckman, X. Wang, K. M. Rabe, D. Vanderbilt, *Phys. Rev. B* **2009**, *79*, 14 144124.

[15] J. Heron, J. Bosse, Q. He, Y. Gao, M. Trassin, L. Ye, J. Clarkson, C. Wang, J. Liu, S. Salahuddin, et al., *Nature* **2014**, *516*, 7531 370.

[16] E. A. Nowadnick, C. J. Fennie, *Phys. Rev. B* **2016**, *94* 104105.

[17] K. Inzani, N. Pokhrel, N. Leclerc, Z. Clemens, S. P. Ramkumar, S. M. Griffin, E. A. Nowadnick, *Phys. Rev. B* **2022**, *105* 054434.

[18] N. A. Benedek, J. M. Rondinelli, H. Djani, P. Ghosez, P. Lightfoot, *Dalton Trans.* **2015**, *44*, 23 10543.

[19] N. A. Benedek, C. J. Fennie, *Phys. Rev. Lett.* **2011**, *106* 107204.

[20] N. A. Benedek, *Inorg. Chem.* **2014**, *53*, 7 3769.

[21] E. Bousquet, M. Dawber, N. Stucki, C. Lichtensteiger, P. Hermet, S. Gariglio, J.-M. Triscone, P. Ghosez, *Nature* **2008**, *452*, 7188 732.

[22] Y. Wang, F.-T. Huang, X. Luo, B. Gao, S.-W. Cheong, *Adv. Mater.* **2017**, *29*, 2 1601288.

[23] X. Xu, Y. Wang, F.-T. Huang, K. Du, E. A. Nowadnick, S.-W. Cheong, *Adv. Funct. Mater.* **2020**, *30*, 42 2003623.

[24] F.-T. Huang, F. Xue, B. Gao, L. Wang, X. Luo, W. Cai, X.-Z. Lu, J. Rondinelli, L. Chen, S.-W. Cheong, *Nat. Commun.* **2016**, *7*, 1 1.

[25] J. Liu, G. Shen, Y. Wang, P. Li, Z. Zhang, X. Chen, F. Yan, X. Chen, H. Shen, J. Zhu, *Ferroelectrics* **1999**, *221*, 1 97.

[26] Y. Ding, J. Liu, I. MacLaren, Y. Wang, *Appl. Phys. Lett.* **2001**, *79*, 7 1015.

[27] D. Su, J. Zhu, Y. Wang, Q. Xu, J. Liu, *J. Appl. Phys.* **2003**, *93*, 8 4784.

[28] Y. Shimakawa, Y. Kubo, Y. Nakagawa, S. Goto, T. Kamiyama, H. Asano, F. Izumi, *Phys. Rev. B* **2000**, *61* 6559.

[29] Ismunandar, B. J. Kennedy, Gunawan, Marsongkohadi, *J. Solid State Chem.* **1996**, *126*, 1 135.

[30] H. T. Stokes, D. M. Hatch, *Isotropy Subgroups of the 230 Crystallographic Space Groups*, WORLD SCIENTIFIC, **1989**.

[31] F.-T. Huang, S.-W. Cheong, *Nat. Rev. Mater.* **2017**, *2*, 3 1.

[32] T. Choi, Y. Horibe, H. Yi, Y. J. Choi, W. Wu, S.-W. Cheong, *Nat. Mater.* **2010**, *9*, 3 253.

[33] S. Artyukhin, K. T. Delaney, N. A. Spaldin, M. Mostovoy, *Nat. Mater.* **2014**, *13* 42.

[34] V. Janovec, *Ferroelectrics* **1981**, *35*, 1 105.

[35] E. Sonin, A. Tagantsev, *Zh. Éksp. Teor. Fiz* **1988**, *67* 396.

[36] E. Sonin, A. Tagantsev, *Ferroelectrics* **1989**, *98*, 1 291.

[37] R. E. Cohen, H. Krakauer, *Phys. Rev. B* **1990**, *42* 6416.

[38] G. Kresse, J. Furthmüller, *Comp. Mater. Sci.* **1996**, *6*, 1 15.

[39] J. P. Perdew, A. Ruzsinszky, G. I. Csonka, O. A. Vydrov, G. E. Scuseria, L. A. Constantin, X. Zhou, K. Burke, *Phys. Rev. Lett.* **2008**, *100* 136406.

[40] G. Henkelman, B. P. Uberuaga, H. Jónsson, *J. Chem. Phys.* **2000**, *113*, 22 9901.

[41] H. T. Stokes, D. M. Hatch, B. J. Campbell, ISOTROPY software suite, URL http://www.stokes.byu.edu/isotropy.html.

[42] B. J. Campbell, H. T. Stokes, D. E. Tanner, D. M. Hatch, *J. Appl. Crystallogr.* **2006**, *39*.

[43] K. Momma, F. Izumi, *J. Appl. Crystallogr.* **2008**, *41*, 3 653.

[44] E. Nowadnick, N. Pokhrel, First-principles calculations on Aurivillius oxides SrBi2(Ta,Nb)2O9, Dryad, Dataset, URL https://doi.org/10.6071/M3XD43.

# Supporting Information

**Ferroelectric Switching Pathways and Domain Structure of SrBi$_2$(Ta,Nb)$_2$O$_9$ from First Principles**

Nabaraj Pokhrel[1], Elizabeth A. Nowadnick[2]

1) Department of Physics, University of California, Merced, Merced, California 95343, USA

2) Department of Materials Science and Engineering, University of California, Merced, Merced, California 95343, USA

# Ferroelectric switching pathways in $SrBi_2Nb_2O_9$

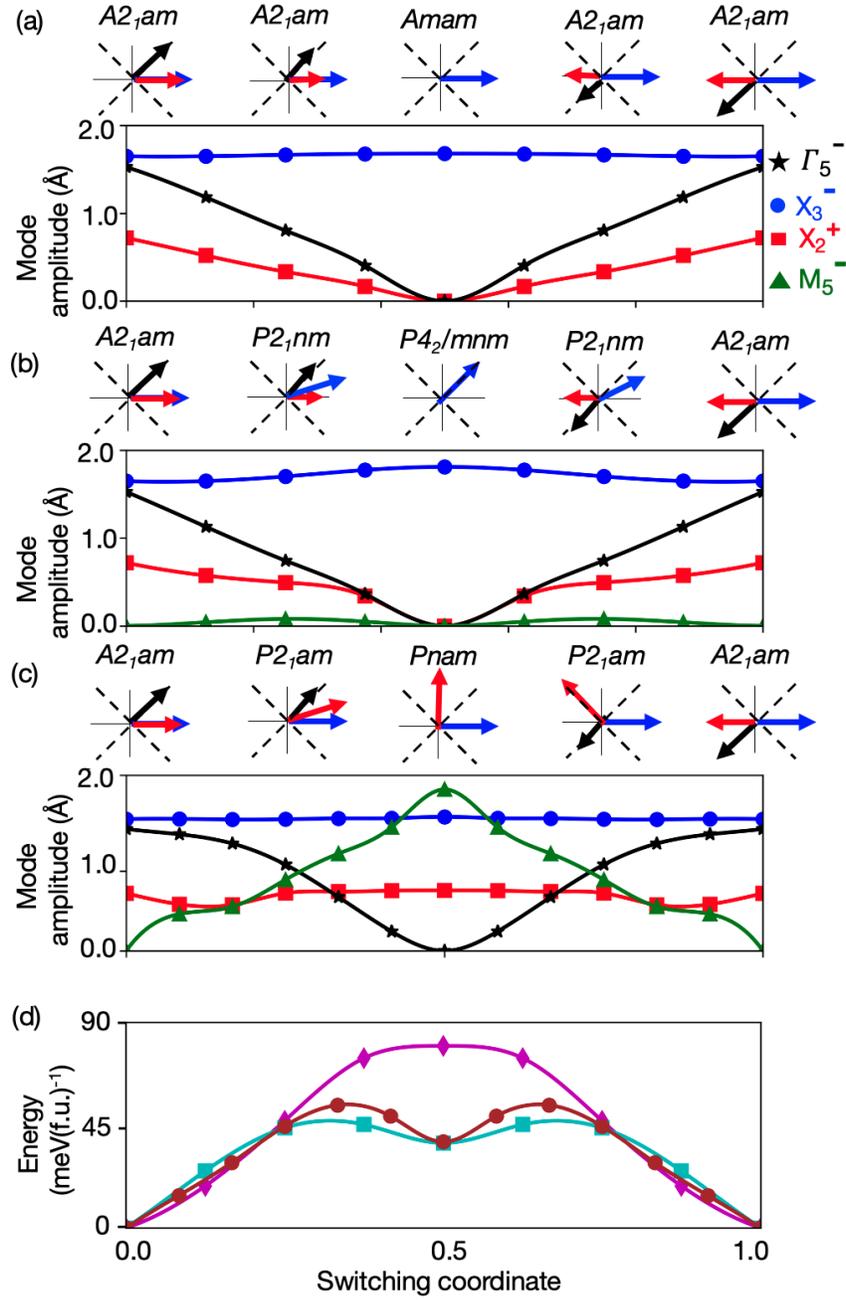

**Supporting Figure S** 1: Ferroelectric switching paths for $SrBi_2Nb_2O_9$: (a) one-step switching, (b) two-step switching via $P4_2/mnm$, and (c) two-step switching via $Pnam$. The upper part of each panel shows how the $X_2^+$ (red), $X_3^-$ (blue), and $\Gamma_5^-$ (black) order parameters evolve during switching, and the bottom part of each panel shows the amplitudes of all order parameters as a function of switching coordinate, obtained from NEB calculations. (d) Total energy as a function of switching coordinate obtained from NEB calculations, for the paths shown in (a)-(c).

# Evolution of octahedral rotations along two-step ferroelectric switching pathways

**Supporting Figure** 2 shows how the crystal structure evolves during the $X_3^-$ and $X_2^+$ order parameter rotations that facilitate the low-energy two-step $P2_1nm$- and $P2_1am$-barrier paths, shown in Figure 3 of the main text. Here, for clarity, the crystal is viewed along the $c$ axis and the two perovskite slabs in the unit cell are colored differently (grey and turquoise). In the $P2_1nm$-barrier path (**Supporting Figure** 2(a)), the $X_3^-$ octahedral tilt axis is initially along [-110]. Then, as the structure enters the $P2_1nm$ phase the $X_3^-$ tilt axis in the grey (upper) perovskite layer starts to rotate clockwise and the tilt axis in the turquoise (lower) layer rotates counterclockwise. Upon reaching $P4_2/mnm$, the octahedra in the grey layer tilt about [100] and the turquoise-layer octahedra tilt about [010], establishing $a^-b^0b^0$ and $b^0a^-b^0$ tilt patterns in the adjacent perovskite layers. Finally, in the second step of the switching process, the tilt axes rotate back to the original [-110] orientation as the $X_2^+$ rotation turns on with the opposite sense. In the $P2_1am$-barrier path (**Supporting Figure** 2(b)), the $X_2^+$ order parameter rotates: in the first step ($A2_1am \rightarrow Pnam$) the octahedral rotation in the turquoise layer turns off and then turns on again with opposite sense upon approaching the $Pnam$ intermediate phase. In the second step ($Pnam \rightarrow A2_1am$) the octahedral rotation in the grey layer turns off and then on again with opposite sense.

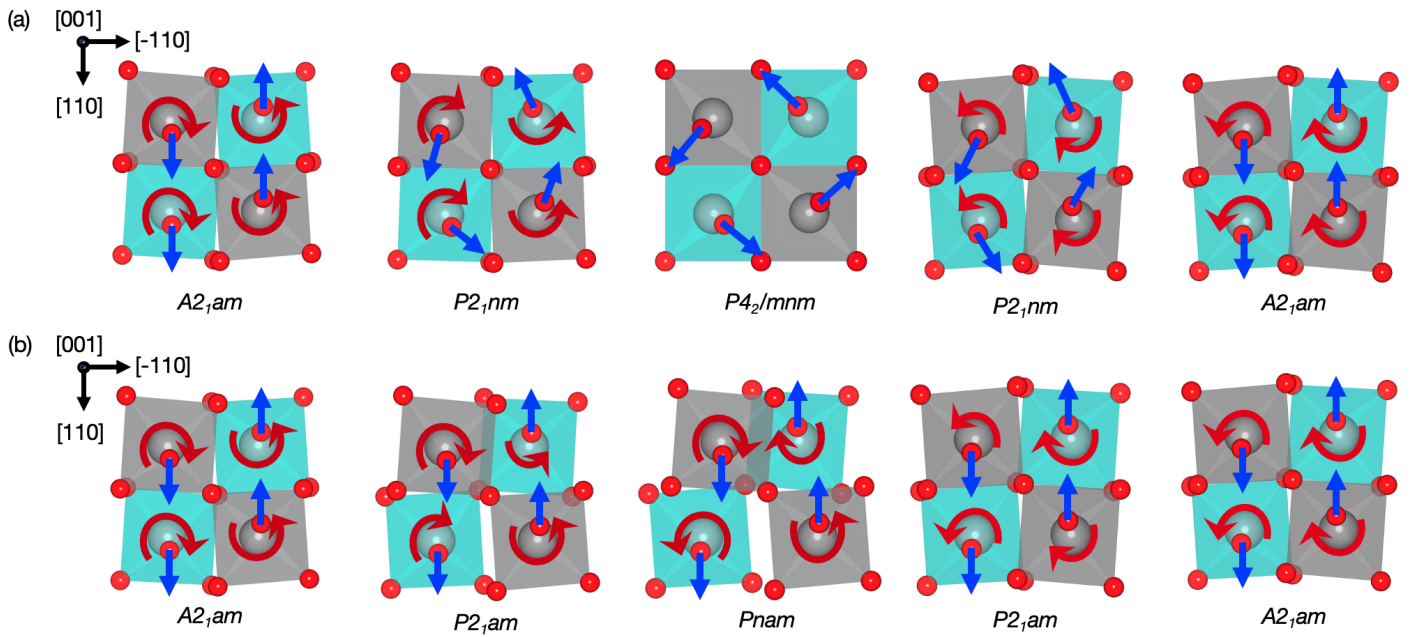

**Supporting Figure S** 2: Evolution of the octahedral rotation and octahedral tilting in $SrBi_2Ta_2O_9$ and $SrBi_2Nb_2O_9$ along the two-step (a) $P2_1nm$-barrier switching pathway, and (b) the $P2_1am$-barrier pathway. Atomic displacements associated with the $X_3^-$ and $X_2^+$ distortions are indicated by blue and red colored arrows, respectively. The upper (lower) perovskite layers in a unit cell are colored grey (turquoise). The Sr and Bi atoms are suppressed for clarity.

# Energy surfaces

This Supporting Note further explores the $P2_1am$-barrier path shown in Figure 3(c) of the main text. In particular, we comment on why the (highest-energy) barrier structure is not found at the switching coordinate midway between the $A2_1am$ ground state and the $Pnam$ intermediate (phase $\phi_{X_2^+} = 45°$) but instead at $\phi_{X_2^+} = 80.5°$. Since going from $A2_1am$ to $Pnam$ involves reversing the $X_2^+$ octahedral rotation in one of the perovskite layers (**Supporting Figure** 2), naively one may expect that the highest energy structure is realized at the midpoint when the octahedral rotation amplitude is zero in one of the layers. However, this does not take into account the key role that the $\Gamma_5^-$ and $M_5^-$ distortions play in the structural energetics. To show this, in **Supporting Figure** 3 we compute energy surfaces by freezing in the $\Gamma_5^-$ and $M_5^-$ distortions into the high-symmetry $I4/mmm$ structure. These energy surfaces show that $I4/mmm$ is unstable with respect to both distortions, but the $\Gamma_5^-$ distortion is significantly more energy lowering than the $M_5^-$ distortion. This is expected, given that the $Pnam$ phase is higher energy than $A2_1am$. These energetics provide a rationalization for why the barrier structure is attained at a point on the switching path closer to the $Pnam$ phase, with a larger $M_5^-$ amplitude and smaller $\Gamma_5^-$.

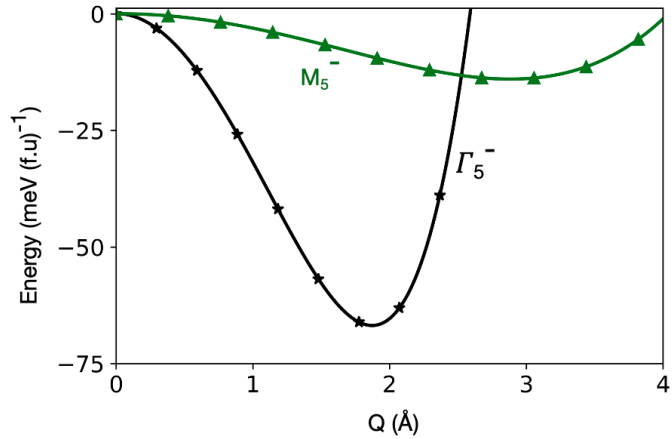

**Supporting Figure S** 3: Energy surface computed by freezing in the $\Gamma_5^-$ (black) and $M_5^-$ (green) distortions into the high-symmetry $I4/mmm$ reference structure of $SrBi_2Ta_2O_9$.

# 90° ferroelectric switching path

In the main text, we considered pathways for 180° ferroelectric switching where the polarization reverses by a full 180°. Ferroelectric switching also can proceed via 90° reorientations of the polarization direction, which is particularly relevant if there are 90° ferroelectric domain walls present in a sample. Interestingly, a small modification of the $P2_1nm$-barrier switching path shown in Figure 3(b) of the main text can describe 90° ferroelectric switching: namely, in the second step the $X_3^-$ order parameter continues to rotate counterclockwise to the $(0,a)$ direction instead of returning to its original orientation. This is shown in **Supporting Figure** 4. The implication of this observation is that, at least at the level of intrinsic ferroelectric switching, 180° and 90° polarization switching have the same energy barrier.

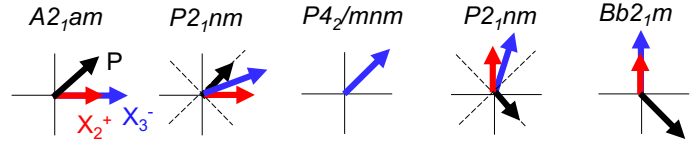

**Supporting Figure S** 4: Example of a 90° ferroelectric switching path between domains in opposite orthorhombic twins.

**Supporting Table S** 1: Distortion amplitudes and lattice parameters of SrBi$_2$Ta$_2$O$_9$ and SrBi$_2$Nb$_2$O$_9$ for all space groups given in Table 3 of the main text. The distortion amplitudes and lattice parameters are given in Å, and the distortion amplitudes are reported in a 56 atom computational cell. For $P2_1am$ and $P2_1nm$, where one of the order parameters is along a low symmetry $(a,b)$ direction, the reported values are those from the maximum energy structure obtained from a NEB calculation.

| Space group | SrBi$_2$Ta$_2$O$_9$ | | | | | | | SrBi$_2$Nb$_2$O$_9$ | | | | | | |
|---|---|---|---|---|---|---|---|---|---|---|---|---|---|---|
| | $Q_{\Gamma_5^-}$ | $Q_{X_2^+}$ | $Q_{X_3^-}$ | $Q_{M_5^-}$ | $a$ | $b$ | $c$ | $Q_{\Gamma_5^-}$ | $Q_{X_2^+}$ | $Q_{X_3^-}$ | $Q_{M_5^-}$ | $a$ | $b$ | $c$ |
| $Imm2$ | 1.75 | - | - | - | 5.547 | 5.547 | 25.073 | 1.76 | - | - | - | 5.554 | 5.554 | 25.010 |
| $F2mm$ | 1.83 | - | - | - | 5.567 | 5.542 | 25.002 | 1.65 | - | - | - | 5.573 | 5.548 | 24.906 |
| $Acam$ | - | 0.91 | - | - | 5.469 | 5.469 | 24.712 | - | 0.92 | - | - | 5.460 | 5.460 | 24.713 |
| $P4/mbm$ | - | 0.55 | - | - | 5.501 | 5.501 | 24.563 | - | 0.81 | - | - | 5.492 | 5.492 | 24.564 |
| $Amam$ | - | - | 1.67 | - | 5.52 | 5.552 | 24.687 | - | - | 1.68 | - | 5.512 | 5.549 | 24.675 |
| $P4_2/mnm$ | - | - | 1.79 | - | 5.541 | 5.541 | 24.580 | - | - | 1.81 | - | 5.538 | 5.538 | 24.566 |
| $A2_1am$ | 1.48 | 0.55 | 1.63 | - | 5.535 | 5.541 | 24.951 | 1.52 | 0.72 | 1.65 | - | 5.543 | 5.541 | 24.891 |
| $Pnam$ | 0 | 0.58 | 1.67 | 1.91 | 5.522 | 5.535 | 24.866 | - | 0.75 | 1.67 | 2.02 | 5.524 | 5.530 | 24.836 |
| $P2_1am$ | 0.56 | 0.52 | 1.65 | 1.21 | 5.520 | 5.539 | 24.805 | 0.67 | 0.74 | 1.65 | 1.22 | 5.517 | 5.531 | 24.779 |
| $P2_1mn$ | 0.50 | 0.21 | 1.72 | 0.11 | 5.533 | 5.546 | 24.661 | 0.36 | 0.34 | 1.77 | 0.07 | 5.532 | 5.541 | 24.599 |
| $Cm2a$ | 1.60 | 0.74 | - | - | 5.536 | 5.541 | 24.956 | 1.61 | 0.96 | - | - | 5.537 | 5.513 | 24.880 |